%% file: main.tex
\documentclass{article}

\usepackage{spconf}

\ninept       

\include{preamble}

\include{ICASSPBibStyle}

\usepackage[skip=5pt]{caption}

\title{Short-and-sparse Deconvolution via Rank-One Constrained Optimization (ROCO)}
\name{Cheng Cheng and Wei Dai}

\address{Department of Electrical and Electronic Engineering, Imperial College London, UK}

\begin{document}

\maketitle

\begin{abstract}
Short-and-sparse deconvolution (SaSD) aims to recover a short kernel and a long and sparse signal from their convolution. 
In the literature, formulations of blind deconvolution is either a convex programming via a matrix lifting of convolution, or a bilinear Lasso. 
Optimization solvers are typically based on bilinear factorizations. 
In this paper, we formulate SaSD as a non-convex optimization with a rank-one matrix constraint, hence referred to as Rank-One Constrained Optimization (ROCO). 
The solver is based on alternating direction method of multipliers (ADMM). 
It operates on the full rank-one matrix rather than bilinear factorizations. 
Closed form updates are derived for the efficiency of ADMM. 
Simulations include both synthetic data and real images. 
Results show substantial improvements in recovery accuracy (at least 19dB in PSNR for real images) and comparable runtime compared with benchmark algorithms based on bilinear factorization. 
\end{abstract}


\section{Introduction}

Blind deconvolution is a problem to recovery two unknown sequences from their convolution. 
This problem arises in many applications, including  
astronomy image \cite{jefferies1993restoration,molina2001image},
remote sensing \cite{jalobeanu2004adaptive,fonseca2009digital},
medical ultrasound image \cite{taxt1999noise,michailovich2007blind} and so on.
Blind deconvolution problems are fundamentally ill-posed as the number of unknown variables is more than the number of observations. 
Extra assumptions (typically in the form of prior knowledge) must be imposed to reduce the number of unknowns. 
Early methods for blind deconvolution include iterative Fourier transform \cite{ayers1988iterative}, maximum likelihood (ML) estimation \cite{katsaggelos1980maximum} and maximum-a-posteriori (MAP) estimation \cite{likas2004variational,fergus2006removing,levin2009understanding,xu2010two}.

In the era of compressive sensing (CS) \cite{donoho2006compressed} and sparse recovery, 
modern techniques including \cite{ahmed2013blind,choudhary2014sparse,ling2015self,chi2016guaranteed,li2016identifiability} assume that either the two unknown sequences lie in fixed low-dimensional spaces, or one is in a fixed low-dimensional space and the other is sparse. 
The seminal work \cite{ahmed2013blind} by Ahmed et. al. studies on the former case and formulates blind deconvolution as a low-rank matrix recovery problem. 
It is based on the observation that a convolution can be viewed as a linear operator acting on a rank-one matrix formed by the outer product of the two unknown sequences. 
By this matrix lifting technique, a convex optimization is then developed where nuclear norm is employed to promote low-rankness. 
As an efficient solver, Burer-Monteiro factorization of low-rank matrices is adopted where the full low-rank matrix is replaced with two factor matrices of much smaller dimensions \cite[Section 2.1]{ahmed2013blind}. 

Recently, short-and-sparse deconvolution (SaSD) receives much attention \cite{zhang2017global,kuo2019geometry,lau2019short,zhang2019structured}. 
It assumes that the convolution is from a short kernel and a sparse signal. 
In stead of matrix lifting, a popular approach \cite{zhang2017global,kuo2019geometry,lau2019short,zhang2019structured} is to cast blind deconvolution as a bilinear Lasso problem and solve it by alternating minimization, where the two unknown sequences are updated alternatively by fixing the other. 
The bilinear Lasso formulation is non-convex. 
Nevertheless, analysis in \cite{zhang2017global,kuo2019geometry} shows that under certain conditions, every local minimum is close to some shift of the ground truth. 
The popularity of bilinear Lasso is based on its good empirical performance and advantages in computational complexity: it has been widely believed that it is expensive to directly operate on the full low-rank matrix in matrix lifting. 

This paper focuses on the SaSD problem. 
Our optimization formulation follows the matrix lifting approach but is with a non-convex constraint that the lifted matrix is of rank-one, hence referred to as Rank-One Constrained Optimization (ROCO). 
An alternating direction method of multipliers (ADMM) algorithm is developed to solve this non-convex optimization problem. 
It works on the full rank-one matrix directly instead of the bilinear Burer-Monteiro factorization. 
To address potential computational complexity issues, closed forms are derived to evaluate the convolution operator and to update the full rank-one matrix in ADMM. 
The numerical tests include both synthetic data and real images from MNIST dataset, and compare ROCO with multiple benchmark algorithms based on bilinear Lasso. 
Simulations show that on one hand ROCO achieves substantial improvements in recovery accuracy (at least 19dB in PSNR for real images), on the other hand the runtime of ROCO is comparable to (sometimes less than) that of benchmark algorithms. 

\section{Background}

The circular convolution of two $N$-point periodic sequences $x[n]$ and $h[n]$ is defined as 
\begin{align*}
    y[n] & = x[n] \circledast h[n] = \sum_{m=0}^{N-1} x[m]h[n-m]. 
\end{align*}
A convolution of two finite-length sequences can be represented by a circular convolution by zero-padding the end of the sequences to a common length $N$. 

Blind deconvolution is to recover both the sequences $x[n]$ and $h[n]$ from their circular convolution $y[n]$. 
It is well known that the solutions to blind deconvolution subject to scaling and shifting ambiguities. 
That is, if $y[n] = x[n] \circledast h[n]$, then $y[n] = (a x[n]) \circledast (h[n] / a)$ 
and $y[n] = S_{\tau} (x[n]) \circledast S_{-\tau}(\bm{h})$,
where $a\ne 0$ is a scaling constant, 
$S_{\tau}(x[n]) = x[n - \tau]$ denotes a shift of the sequence $x[n]$ by $\tau$ positions,
and $S_{-\tau}(h[n]) = h[n + \tau]$ denotes a shift in the other direction. 

This paper focuses on the problem of short-and-sparse blind deconvolution \cite{kuo2019geometry,lau2019short}. 
Consider a convolution of two finite-length sequences: 
without loss of generality assume that $x[n]$ is sparse and $h[n]$ is short. 
Write the sequences as vectors, i.e., $\bm{x},\bm{y} \in \mathbb{R}^N$ and $\bm{h} \in \mathbb{R}^M$ with $M \ll N$. 
Blind deconvolution can be written as 
\begin{equation}
    \min_{\bm{x},\bm{h}}~ \Vert \bm{x} \Vert_{0} \quad \text{s.t.}~ \bm{y}=\bm{x} \circledast \bm{h},
    \label{eq:SaSD-L0}
\end{equation}
where $\Vert \cdot \Vert_0$ denotes $\ell_0$ pseudo-norm which counts the number of non-zero elements.
As a convex relaxation of the sparsity,  the $\ell_0$ pseudo-norm in \eqref{eq:SaSD-L0} can be replaced by the $\ell_1$-norm. 
To partially address the scaling ambiguity, an extra constraint $\Vert \bm{h} \Vert_2 = 1$ can be added. 
These yield the well-known bilinear Lasso form \cite{lau2019short}
\begin{equation}
    \min_{\bm{x},\bm{h}}~ 
    \frac{1}{2}\Vert\bm{y}-\bm{x} \circledast \bm{h} \Vert_{2}^{2}
    +\lambda\Vert\bm{x}\Vert_1, \quad
    \text{s.t.}~ \Vert\bm{h}\Vert_2=1.
    \label{eq:SaSD-L1}
\end{equation}

The work \cite{lau2019short} summarizes several popular methods to solve \eqref{eq:SaSD-L1}.
The basic approach is the Alternating Descent Method (ADM) where in each iteration, one updates $\bm{x}$ by fixing $\bm{h}$ and then alternatively updates $\bm{h}$ (using a Riemannian gradient descent \cite{absil2009optimization}) by fixing $\bm{x}$. 
It can be shown that the objective function decreases monotonically and hence a convergence is guaranteed. 
To speed up the convergence, two other variations are studied in \cite{lau2019short}: the inertial Alternating Descent Method (iADM) adds momentum to the descent direction to mitigate possible oscillations in the optimization process, and the homotopy-ADM method applies a homotopy continuation method on top of ADM.

The optimization \eqref{eq:SaSD-L1} is still non-convex due to the bilinear term in the objective function. 
On the other hand, SaSD problem \eqref{eq:SaSD-L0} can be relaxed into a convex optimization problem by using the matrix lifting technique \cite{ahmed2013blind}. 
This is based on the observations that $\bm{Z} := \bm{x}\bm{h}^{\mathsf{T}}$ is a rank-one matrix, that the convolution can be written as a linear operator acting on $\bm{Z}$, i.e., $\bm{y} = \mathcal{A}(\bm{Z})$, and that the sparsity in $\bm{x}$ can be translated into row sparsity of $\bm{Z}$. 
Hence, SaSD problem can be reformulated as  
\begin{align}
    \min_{\bm{Z}}~
    & \Vert \bm{Z} \Vert_* + \lambda_1 \Vert \bm{Z}^{\mathsf{T}} \Vert_{2,1} + \frac{\lambda_2}{2} \Vert \bm{y} - \mathcal{A}(\bm{Z}) \Vert_2^2, 
    \label{eq:SaSD-lifting-convex}
\end{align}
where the nuclear norm $\Vert\cdot\Vert_*$ promotes low-rank solutions and the $\ell_{2,1}$-norm $\Vert \cdot \Vert_{2,1}$ promotes a solution $\bm{Z}$ with a small number of non-zero rows. 
However, there is not much discussion of directly solving \eqref{eq:SaSD-lifting-convex} in the literature. This is partly because the performance of \eqref{eq:SaSD-lifting-convex} is typically not as good as that of the non-convex counterpart \eqref{eq:SaSD-L1}, partly due to the widespread belief that it is much more computationally efficient to operate on the low-rank Burer-Monteiro factorization rather than the full matrix directly \cite{zhang2017global,ling2019regularized}.

\section{SaSD via ROCO}

Our approach to address SaSD problem is based on the matrix lifting technique in \eqref{eq:SaSD-lifting-convex} but uses a non-convex optimization formulation. 

\subsection{Optimization Formulation}

Consider the rank-one matrix lifting of a convolution \cite{ahmed2013blind,choudhary2014sparse}. In particular, 
\begin{align}
    \bm{y} 
    & = \bm{x} \circledast \bm{h} 
    = \mathcal{A} \left( \bm{x} \bm{h}^{\mathsf{T}} \right) 
    \nonumber \\
    & = \mathcal{A}(\bm{Z}) 
    = \bm{A} \text{vec}(\bm{Z}), 
\end{align}
where $\bm{Z} := \bm{x} \bm{h}^{\mathsf{T}}$ is clearly a rank-one matrix, the linear operator $\mathcal{A}:~\mathbb{R}^{N \times M} \rightarrow \mathbb{R}^{N}$ can be represented by the matrix 
\begin{align}
    \bm{A} 
    & := \left[ \bm{I}_N, \mathcal{S}_1(\bm{I}_N), \cdots, \mathcal{S}_{M-1}(\bm{I}_N) \right]
    \in \mathbb{R}^{N \times MN},
\end{align}
$\bm{I}_N$ is the $N \times N$ identity matrix, and $\mathcal{S}_\tau (\cdot)$ is the \emph{downwards} circular shift operator that circularly moves the rows of the input vector/matrix downwards by $\tau$ many positions. 

Our optimization formulation of blind deconvolution is based on the matrix lifting technique in \eqref{eq:SaSD-lifting-convex}. 
Suppose that $\bm{x}$ is sparse. 
A zero entry in $\bm{x}$, say $x_n$, leads to a zero row in $\bm{Z}$, i.e., $\bm{Z}_{n,:} = x_n \bm{h}^{\mathsf{T}} = \bm{0}^{\mathsf{T}}$. 
Hence, to promote sparsity in $\bm{x}$ is equivalent to enforcing row-sparsity in the matrix $\bm{Z}$. 
Define row-wise $\ell_{2,0}$-norm as 
\begin{align}
    \Vert \bm{Z} \Vert _{r,2,0}
    & :=\left\Vert \left[
        \left\Vert\bm{Z}_{1,:}\right\Vert_{2}, \left\Vert\bm{Z}_{2,:}\right\Vert_{2},
        \cdots,\left\Vert\bm{Z}_{N,:}\right\Vert_{2}
    \right]^{\mathsf{T}} \right\Vert_{0},
    \label{eq:L-row-20}
\end{align}
which counts the number of non-zero rows of the input matrix $\bm{Z}$. Further define the set of rank-one matrices as 
\begin{align}
    \mathcal{R}1 
    & = \left\{ \bm{Z}\in\mathbb{R}^{N\times M}:~ \text{rank} \left(\bm{Z}\right)\leq 1 \right\}. 
\end{align}
Then the SaSD problem can be formulated as 
\begin{align}
    \min_{\bm{Z}}~ 
     \Vert \bm{Z} \Vert _{r,2,0}, 
\quad
    \text{s.t.}~
     \bm{y}= \bm{A}\text{vec}(\bm{Z}),~ 
    \bm{Z}\in\mathcal{R}1.
\end{align}
Further relax the non-convex $\ell_0$ pseudo-norm into the convex $\ell_1$-norm, yielding 
\begin{align}
    \min_{\bm{Z}}~ 
     \Vert \bm{Z} \Vert _{r,2,1},
    \label{eq:SaSD-Opt-Formulation}
    \quad
    \text{s.t.}~
   \bm{y}= \bm{A}\text{vec}(\bm{Z}),~ 
    \bm{Z}\in\mathcal{R}1,
\end{align}
where $\Vert \cdot \Vert_{r,2,1}$ is defined by replacing the $\ell_0$ pseudo-norm in \eqref{eq:L-row-20} by $\ell_1$-nrom. 
It is important to note the non-convex rank constraint in \eqref{eq:SaSD-Opt-Formulation}, which marks the key difference between our approach and the convex counterpart \eqref{eq:SaSD-lifting-convex}. 

The convolution operator $\bm{y}= \mathcal{A}(\bm{Z}) = \bm{A}\text{vec}(\bm{Z})$ can be evaluated efficiently. The straightforward way is to write the matrix $\bm{A}$ explicitly, store it using a sparse matrix data type (for example in Matlab), and then compute $\bm{A}\text{vec}(\bm{Z})$. By contrast, we calculate $\bm{A}\text{vec}(\bm{Z})$ by introducing an auxiliary matrix 
\begin{align*}
    \bm{Z}_{\text{CL}} 
    & = \mathcal{S}_\text{CL}(\bm{Z})
    := \left[ \bm{Z}_{:,1}, \mathcal{S}_1(\bm{Z}_{:,2}),\cdots, \mathcal{S}_{M-1}(\bm{Z}_{:,M}) \right],
\end{align*}
where the subscript $\text{CL}$ stands for \emph{Cyclic Lifting}, meaning that $\bm{Z}_{\text{CL}}$ is not a direct matrix lifting $\bm{x}\bm{h}^{\mathsf{T}}$ but a cyclic shifted version of it. Then the convolution can be computed by summing the elements in each row of the auxiliary matrix $\bm{Z}_{\text{CL}}$, i.e., 
\begin{align}
    \bm{y} = & \bm{x} \circledast \bm{h} = \mathcal{A}(\bm{Z}) := \mathcal{S}_\text{CL}(\bm{Z}) \bm{1} = \bm{Z}_{\text{CL}} \bm{1},
    \label{eq:Convolution-A-Operator}
\end{align}
where $\bm{1} \in \mathbb{R}^M$ is the vector of which all elements are one. 
This method minimizes the computation and storage costs of evaluating $\mathcal{A}(\bm{Z})$, and also leads to the efficient update \eqref{eq:ADMM-Z-Update-Closed-Form} of the ADMM solver developed in Section \ref{sub:ADMM-Solver}. 

An ADMM solver is developed to solve \eqref{eq:SaSD-Opt-Formulation} in Section \ref{sub:ADMM-Solver}. Note the developed ADMM operates on the full matrix $\bm{Z} \in \mathbb{R}^{N \times M}$. By the shortness of $\bm{h}$, $M \ll N$. Each iteration of ADMM has complexity $O(MN^2)$, which is comparable to that of iteration in methods based on sparse Lasso when $M=O(1)$.  

After solving \eqref{eq:SaSD-Opt-Formulation}, SVD is used to extract $\bm{x}$ and $\bm{h}$ from $\bm{Z}$. Consider the SVD $\bm{Z} = \sigma \bm{u} \bm{v}^{\mathsf{T}}$ (as $\bm{Z}$ has rank one). Let $s \in \{-1,1\}$ be the sign of the first nonzero entry in the vector $\bm{v}$. We set $\bm{x} = s \sigma \bm{u}$ and $\bm{h} = s \bm{v}$. 

\subsection{An ADMM solver for ROCO}
\label{sub:ADMM-Solver}

We solve the ROCO optimization problem \eqref{eq:SaSD-Opt-Formulation} using an ADMM algorithm. As will be shown soon, each step of ADMM iterations admits a closed form solution, and hence the overall ADMM algorithm is computationally efficient. 

The ADMM reformulation of \eqref{eq:SaSD-Opt-Formulation} is given by 
\begin{align}
    \min_{\bm{P},\bm{Q},\bm{Z}}~ 
    & \Vert \bm{P}\Vert _{r,2,1} 
    + \mathbbm{1}_{\text{rank}(\cdot) \le 1}(\bm{Q})
    \label{eq:SaSD-ROCO-ADMM} \\
    \text{s.t.}~ 
    & \bm{y}=\mathcal{S}_{\text{CL}}(\bm{Z})\bm{1},~
    \bm{P}=\bm{Z},~
    \bm{Q}=\bm{Z},
    \nonumber
\end{align}
where 
\begin{equation*}
    \mathbbm{1}_{\text{rank}(\cdot) \le 1} (\bm{Q})
    :=\begin{cases}
        0, & \text{if}~ \text{rank}(\bm{Z}) \le 1, \\
        +\infty, & \text{otherwise},
    \end{cases} 
\end{equation*}
is the indicator function of a matrix of which the rank is at most one. 
The corresponding augmented Lagrangian can be written as
\begin{align}
    \mathcal{L}_{\rho}
    & =\Vert\bm{P}\Vert_{2,1} 
    + \mathbbm{1}_{\text{rank}(\cdot) \le 1}(\bm{Q}) + \frac{\rho}{2}\Vert\bm{Z}-\bm{P}+\bm{\Lambda}_{1}\Vert_{F}^{2}  
    \nonumber \\
    & 
    + \frac{\rho}{2}\Vert\bm{Z}-\bm{Q}+\bm{\Lambda}_{2}\Vert_{F}^{2}+\frac{\rho}{2}\Vert\mathcal{S}_{CL}(\bm{Z})\bm{1}-\bm{y}+\bm{\lambda}_{0}\Vert_{2}^{2} 
    \nonumber \\
    &-\frac{\rho}{2}\Vert\bm{\Lambda}_{1}\Vert_{F}^{2}
    -\frac{\rho}{2}\Vert\bm{\Lambda}_{2}\Vert_{F}^{2}
    -\frac{\rho}{2}\Vert\bm{\lambda}_{0}\Vert_{2}^{2},
    \nonumber
\end{align}
where $\rho>0$ denotes the penalty parameter, and $\bm{\Lambda}_1 \in \mathbb{R}^{N \times M}$, $\bm{\Lambda}_2 \in \mathbb{R}^{N \times M}$, and $\bm{\lambda}_0 \in \mathbb{R}^{N}$ are \emph{normalized} Lagrange multipliers (for the purpose of notational simplification). 
Hence the ADMM iterations are given by
\begin{align}
    \bm{P}^{l+1}
    & = \arg~ \min_{\bm{P}}~
    \Vert \bm{P} \Vert_{r,2,1}
    +\frac{\rho}{2} \Vert \bm{Z}^{l}-\bm{P}+\bm{\Lambda}_{1}^{l}\Vert _{F}^{2},
    \label{eq:ADMM-P-update} \\
    \bm{Q}^{l+1}
    & = \arg~ \min_{\bm{Q}}~
    \mathbbm{1}_{\text{rank}(\cdot) \le 1}(\bm{Q}) 
    + \frac{\rho}{2} \Vert  \bm{Z}^{l}-\bm{Q} + \bm{\Lambda}_{2}^{l} \Vert _{F}^{2},
    \label{eq:ADMM-Q-update} \\
    \bm{Z}^{l+1} 
    & = \arg~ \min_{\bm{Z}}~ 
    \Vert \bm{Z}-\bm{P}^{l+1}+\bm{\Lambda}_{1}^{l} \Vert_{F}^{2} 
    + \Vert \bm{Z}-\bm{Q}^{l+1}+\bm{\Lambda}_{2}^{l} \Vert_{F}^{2} \nonumber \\  
    & \quad + \Vert \mathcal{S}_{\text{CL}} (\bm{Z})\bm{1} - \bm{y} + \bm{\lambda}_{0}^{l} \Vert_{2}^{2}
    \label{eq:ADMM-Z-update} \\
    \bm{\Lambda}_{1}^{l+1} 
    & = \bm{\Lambda}_{1}^{l}+\bm{Z}^{l+1}-\bm{P}^{l+1},
    \label{eq:ADMM-Lambda1-Update} \\
    \bm{\Lambda}_{2}^{l+1} 
    & =\bm{\Lambda}_{2}^{l}+\bm{Z}^{l+1}-\bm{Q}^{l+1},
    \label{eq:ADMM-Lambda2-Update}\\
    \bm{\lambda}_{0}^{l+1} 
    & = \bm{\lambda}_{0}^{l} + \mathcal{S}_{CL}(\bm{Z})\bm{1} -\bm{y},
    \label{eq:ADMM-Lambda0-Update}
\end{align}
where $l$ is the index of iterations. 

All the three sub-problems \eqref{eq:ADMM-P-update}-\eqref{eq:ADMM-Z-update} involved in ADMM iterations admit closed form solutions. 
In particular, the closed form solution of \eqref{eq:ADMM-P-update} can be obtained by setting the sub-gradient of the objective function to zero. Define $\hat{\bm{P}} := \bm{Z}^{l}+\bm{\Lambda}_{1}^{l}$. Then the solution $\bm{P}^{l+1}$ to \eqref{eq:ADMM-P-update} is given by 
\begin{equation*}
    \bm{P}^{l+1}_{n,:}
    = \left(1-\frac{1}{\rho\Vert\hat{\bm{P}}_{n,:}\Vert_{2}}\right)_{+} \hat{\bm{P}}_{n,:},
\end{equation*}
where $(x)_+ := \max(0,x)$. 

Though the sub-problem \eqref{eq:ADMM-Q-update} is non-convex, its optimal solution can be obtained based on Eckart-Young-Mirsky theorem \cite{golub1987generalization}. 
Define $\hat{\bm{Q}} = \bm{Z}^{l}+\bm{\Lambda}_{2}^{l}$. 
Consider the Singular Value Decomposition (SVD) $\hat{\bm{Q}}=\bm{U} \text{diag}(\bm{\sigma}) \bm{V}^{\mathsf{T}}$. 
The optimal solution of \eqref{eq:ADMM-Q-update} is then given by 
\begin{equation*}
    \bm{Q}^{l+1}
    = \bm{\sigma}_1\bm{U}_{:,1}\bm{V}_{:,1}^T. 
\end{equation*}

The sub-problem \eqref{eq:ADMM-Z-update} is a quadratic optimization, of which the optimal solution can be obtained via pseudo-inverse or conjugate gradient method \cite{nocedal2006conjugate} in principle. 
On the other hand, the specific structure of this quadratic optimization problem allows a much more efficient way to compute its optimal solution. 
Define 
$\bm{C}_1^l := \bm{P}^{l+1}-\bm{\Lambda}_{1}^{l}$, 
$\bm{C}_2^l := \bm{Q}^{l+1}-\bm{\Lambda}_{2}^{l}$,  
$\bm{c}_3^l := \bm{y}-\bm{\lambda}_{0}^{l}$, and 
$\bm{C}_3^l := \left[ \bm{c}_3^l, \cdots, \bm{c}_3^l \right] \in \mathbb{R}^{N \times M}$ ($M$ repeated columns). 
Define $\mathcal{S}_{-\tau} (\cdot)$ is the \emph{upwards} circular shift operator that circularly moves the rows of the input vector/matrix upwards by $\tau$ many positions. Define 
\begin{align*}
  \mathcal{S}_\text{-CL}(\bm{C})
    = \left[ \bm{C}_{:,1}, \mathcal{S}_{-1}(\bm{C}_{:,2}),\cdots, \mathcal{S}_{-(M-1)}(\bm{C}_{:,M}) \right]. 
\end{align*}
It can be verified by linear algebra and Woodbury matrix identity that the optimal solution to \eqref{eq:ADMM-Z-update} is given by 
\begin{align}
    \bm{Z}^{l+1}
    & = \frac{1}{2} \mathcal{S}_{\text{-CL}} \left(
        \left( \mathcal{S}_{\text{CL}}(\bm{C}_{1}^{l}+\bm{C}_{2}^{l}) + \bm{C}_{3}^{l} \right)
        \left( \bm{I}_{M} - \frac{1}{M+2} \bm{1} \bm{1}^{\mathsf{T}} \right) 
    \right).
    \label{eq:ADMM-Z-Update-Closed-Form}
\end{align}

It is noteworthy that problem \eqref{eq:SaSD-ROCO-ADMM} that ADMM is applied to is non-convex. Generally speaking, non-convex ADMM algorithm may not converge. Nevertheless, the above developed ADMM algorithm with sufficient large $\rho$ converges in all of our numerical tests. Interested readers may refer to \cite{wang2019global} for a recent work on adapting non-convex and non-smooth ADMM so that global convergence is guaranteed. 

\begin{figure}[b]
    \captionsetup[subfloat]{farskip=0pt,captionskip=1pt}

    \begin{centering}

    \subfloat[ROCO]{\begin{centering}
    \includegraphics[scale=0.28]{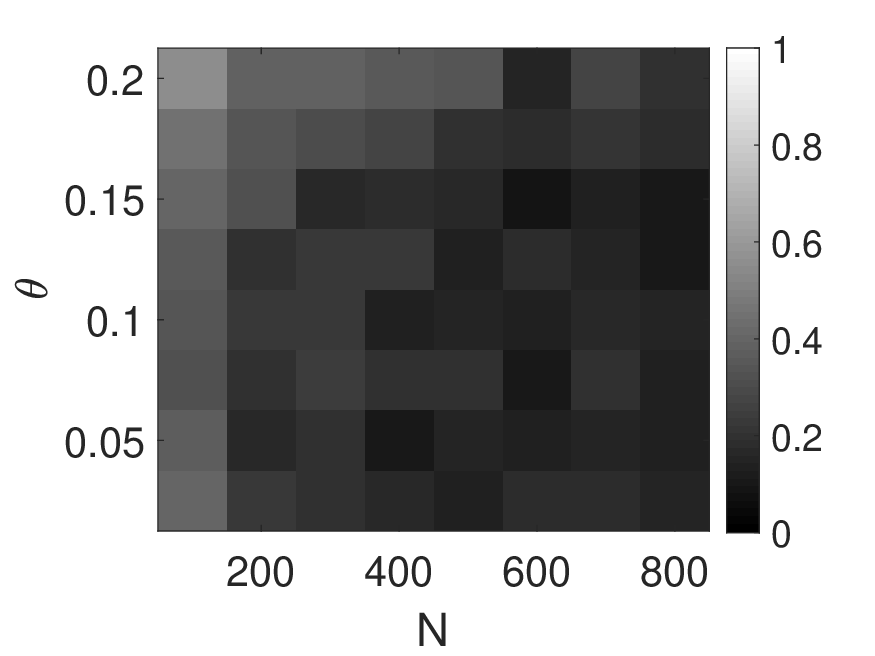}
    \par\end{centering}
    }

    \subfloat[ADM]{\begin{centering}
    \includegraphics[scale=0.28]{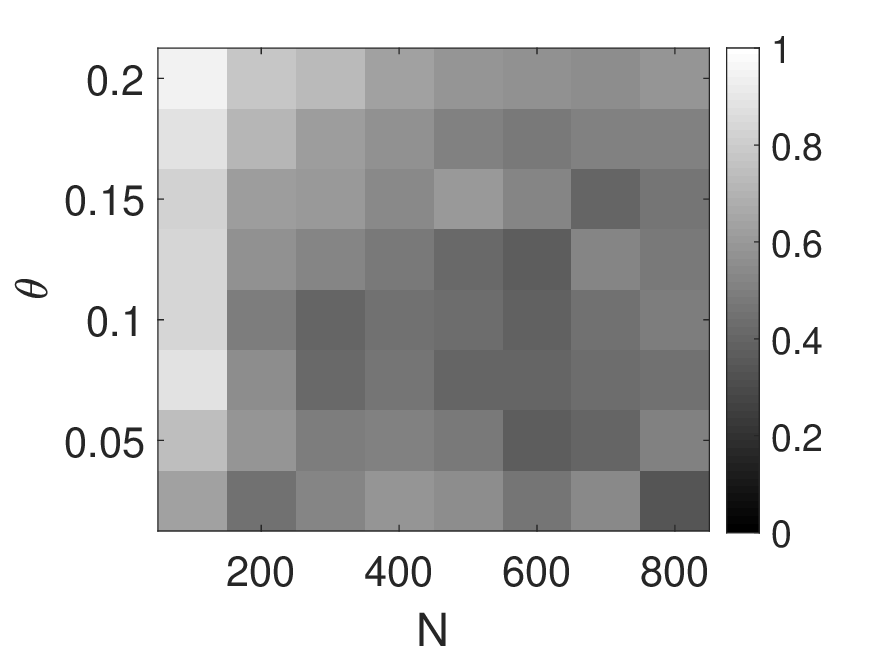}
    \par\end{centering}
    }\subfloat[iADM]{\begin{centering}
    \includegraphics[scale=0.28]{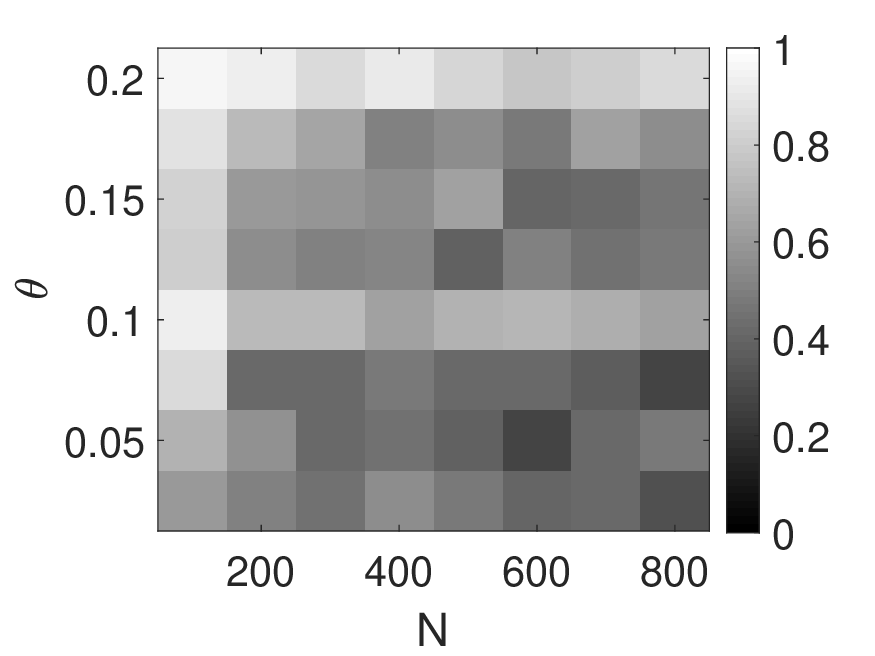}
    \par\end{centering}
    }
    \par\end{centering}
    \begin{centering}
    \subfloat[homotopy-ADM]{\begin{centering}
    \includegraphics[scale=0.28]{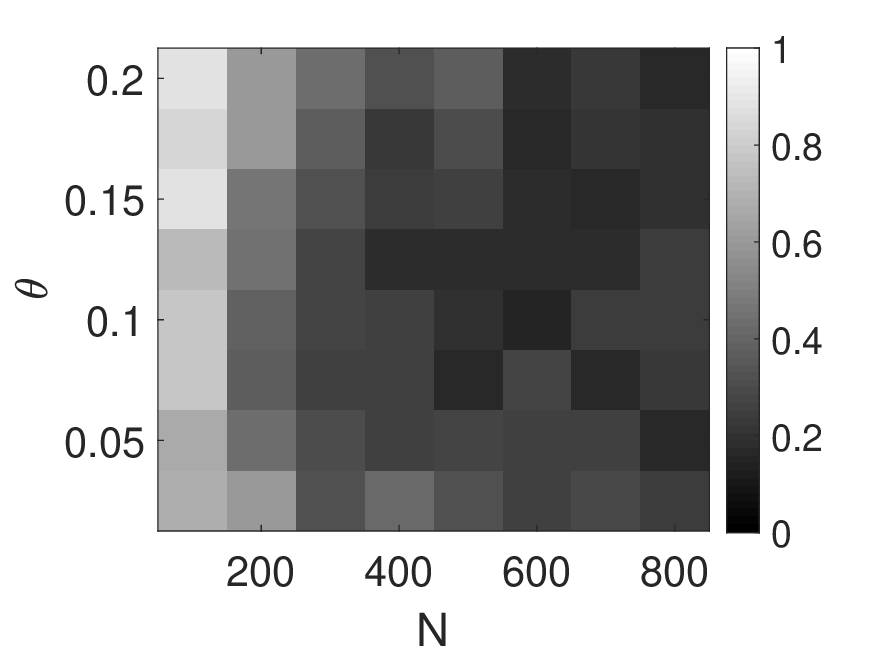}
    \par\end{centering}
    }\subfloat[homotopy-iADM]{\begin{centering}
    \includegraphics[scale=0.28]{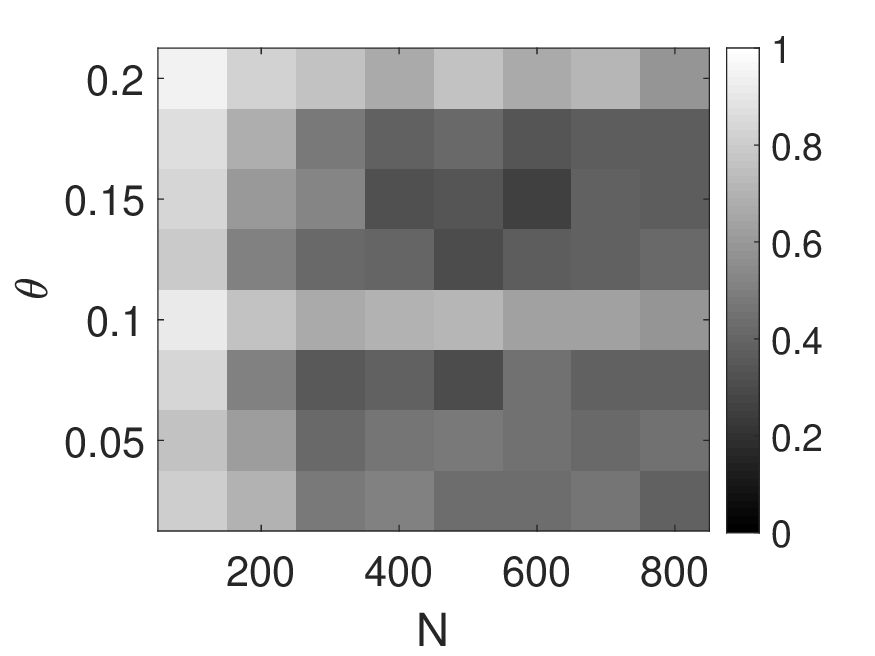}
    \par\end{centering}
    }

    \par\end{centering}
    \caption{\label{fig:Chap5-SaSD-syn-t1}Comparison of SaSD methods based on the failure rate of reconstructing $\bm{h}$. $M=10$, $N=[100,800]$, $\theta=[0.025,0.2]$.}
\end{figure}

\begin{figure}[b]
    \centering
    \includegraphics[scale=0.35]{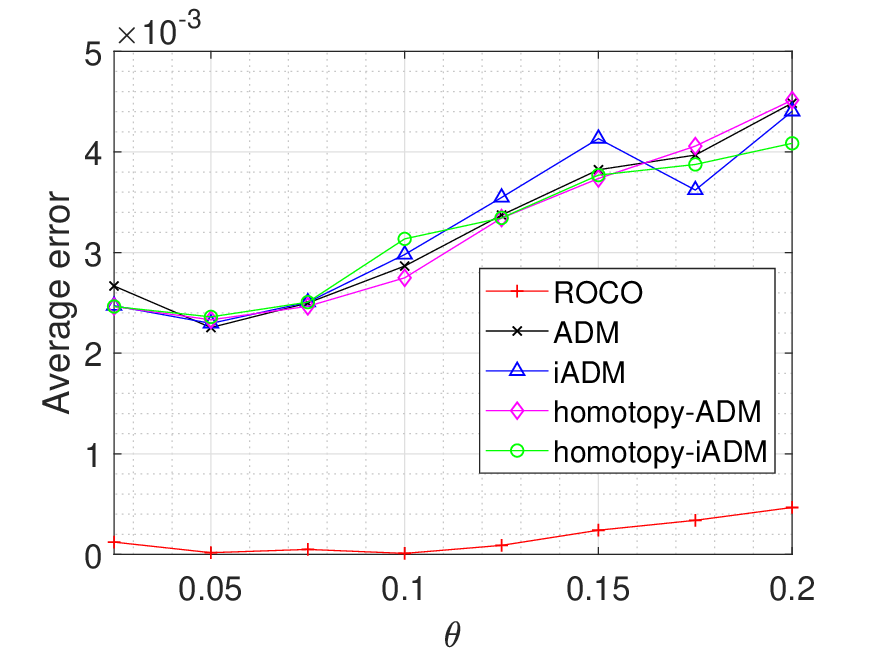}
    \caption{Comparison of SaSD methods based on the average error of successful reconstruction of $\bm{h}$. $M=10$, $N=800$, $\theta=[0.025,0.2]$.}
    \label{fig:Chap5-SaSD-syn-t4}
\end{figure}

\begin{table*}

    \caption{\label{tab:image-deblur}Comparison of different SaSD algorithms applied to image deblurring. Note that the PSNR (in dB) values for the benchmark algorithms are computed using the shifts of their results.}

    \renewcommand{\arraystretch}{1.2}
    \begin{centering}
    \begin{tabular}{|C{2cm}|C{2cm}|C{2cm}|C{2cm}|C{2cm}|C{2cm}|C{2cm}|}
        \hline
        Ground truth digits & Blurred digits & ROCO & ADM & iADM & homotopy-ADM & homotopy-iADM \\ 

        \includegraphics[scale=0.63]{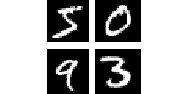} &
        \includegraphics[scale=0.63]{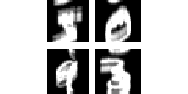} &
        \includegraphics[scale=0.63]{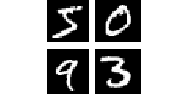}  & \includegraphics[scale=0.63]{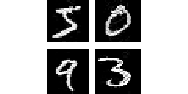} & 
        \includegraphics[scale=0.63]{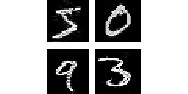} &
        \includegraphics[scale=0.63]{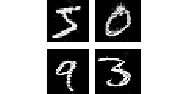} & \includegraphics[scale=0.63]{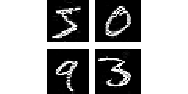} \\

        \hline
        \multicolumn{2}{|c|}{PSNR for digit 5 (dB)}
        & 89.0427 & 20.3754 & 15.4896 & 20.0855 & 15.1360 \\
        \hline
        \multicolumn{2}{|c|}{PSNR for digit 0 (dB)} & 38.6332 & 17.8752 & 19.0489 & 16.4804 & 14.7691 \\
        \hline
        \multicolumn{2}{|c|}{PSNR for digit 9 (dB)} & 51.4815 & 22.8923 & 15.5526 & 16.1069 & 15.7989 \\
        \hline
        \multicolumn{2}{|c|}{PSNR for digit 3 (dB)} & 42.1422 & 13.5715 & 12.6990 & 12.7768 & 12.6003 \\
        \hline

        \multicolumn{2}{|c|}{Average running time (s)} & 2.198 & 1.836 & 1.565 & 2.776 & 2.978 \\

        \hline
    \end{tabular}
    \par\end{centering}
\end{table*}

\section{Numerical tests}  \label{sec:Chap5-numerical-tests}

We compare the ROCO method with four benchmark algorithms: ADM, iADM, homotopy-ADM and homotopy-iADM  \cite{lau2019short}.

\subsection{Synthetic data tests}

The setup for synthetic data tests are specified as follows. 
We assume that the observed data $\bm{y}\in\mathbb{R}^{N}$ is generated from a convolution of  a ground-truth sparse signal $\bm{x}^0\in\mathbb{R}^{N}$ and a ground-truth kernel $\bm{h}^0\in\mathbb{R}^{M}$ via $\bm{y}=\bm{x}^0 \circledast  \bm{h}^0$.
The kernel $\bm{h}^0$ is generated by first filling it with independent realizations of the standard Gaussian variable and then normalizing it to have unit $\ell_2$-norm. 
The sparse vector $\bm{x}^0$ is generated by Bernoulli-Gaussian distribution $\mathcal{BG}(\theta)$, i.e. the occurrence of the non-zero entries in $\bm{x}^0$ is under Bernoulli distribution with probability $\theta$, and the value of non-zeros are under Gaussian distribution. 
In particular, we set the dimension of the kernel $\bm{h}^{0}$ as $M=10$ and vary the sparsity ratio $\theta$ and number of samples $N$ from $0.025$ to $0.2$ and from $100$ to $800$ respectively. The tests are repeated for 100 random trials. 
For ROCO, we set the penalty parameter $\rho=2$.
For all tested algorithms, we stop the optimization process either at 1000 iterations, or
when the difference between two consecutive iterations in $\ell_2$-norms is smaller than threshold $10^{-6}$.

Performance metric includes recovery failure rate and average $\ell_2$-error for success recoveries. 
Denote $\hat{\bm{h}}$ and $\bm{h}^{0}$ as the estimated kernel and the ground-truth kernel respectively. 
The recovery error is defined as 
\begin{align*}
    e_{\text{Rec}}=
    & \min~ \{
        \Vert s_{\tau}(\hat{\bm{h}})-\bm{h}^0\Vert_2, 
        \Vert s_{\tau}(\hat{\bm{h}})+\bm{h}^0\Vert_2:~
        \tau \in [N-1]
        \}
\end{align*}
to address both sign and shift ambiguities. 
Let $e_{\text{Thr}} = 10^{-2}$. 
We call a trial with $e_{\text{Rec}} \le e_{\text{Thr}}$ a successful recovery and otherwise a failed recovery. 
The failure recovery rate is depicted in Figure \ref{fig:Chap5-SaSD-syn-t1}, where the number of trials is 100. 
A darker colour means a lower failure rate. 
It is clear that ROCO has lower failure rate in the tested range of sparsity ratios $\theta$ and signal lengths $N$. 
The improvement of ROCO becomes more significant when the signal length $N$ is relatively small. 

We are also interested in the average error of successful recoveries, as it gives more detailed performance information. 
We focus on the case where $N=800$. 
Results are presented in Figure \ref{fig:Chap5-SaSD-syn-t4}. 
It is clear that ROCO substantially outperforms all other benchmark algorithms. 
Even for less sparse signals $\bm{x}$ where $\theta=0.2$, the average error of successful recoveries is approximately only $1/10$ of that of other algorithms.

\subsection{Image deblurring}

This test is based on real images from MNIST dataset. 
MNIST dataset contains images of handwritten digits from 0 to 9, each of which is of the size $28 \times 28$ pixels.
We randomly choose 4 images (corresponding to 4 different digits) from MNIST dataset as ground truth images, denoted as $\bm{I}^{0}$.
We also generate a convolutional kernel $\bm{h}^{0} \in \mathbb{R}^9$ using the same way as in the synthetic data test. 
The observation $\bm{y}$ is a blurred image generated by a convolution of $\text{vec}(\bm{I}^0)$ and $\bm{h}^0$. 
The SaSD problem is then a single image deblurring problem. 
In simulations, we set $\rho=20$ for ROCO method. 

Table \ref{tab:image-deblur} summarises the simulation results. 
The performance metric is peak signal-to-noise ratio (PSNR) defined as 
\begin{equation*}
    \text{PSNR} 
    =\max \left\{
        10{\rm log}_{10}\frac{N}{\Vert S_{\tau}(\hat{\bm{I}})-\bm{I}^0\Vert_{F}^{2}}:~
        \forall \tau\in [\sqrt{N}-1]
    \right\},
\end{equation*}
where $N = 28 \times 28$, $\sqrt{N}-1 = 27$, and the term $S_{\tau}$ is introduced to address shift ambiguity. 
From the results, the recovered images via ROCO are visually much sharper than those from other methods, and actually look identical to the ground truth ones. 
In terms of PSNR, the performance of ROCO is at least 19dB better than those of other methods. 
It is important to note that the run time of ROCO is comparable to benchmark algorithms (sometimes less). 
This shows that matrix lifting may not sacrifice in computational complexity compared with factorization based methods, which is against the widespread wisdom in the literature.

\section{Conclusion}

In this paper, we develop the ROCO method for the SaSD problem. 
The distinct characteristic is that ROCO operates on full rank-one matrices in both formulation and its ADMM solver. 
For an efficient ADMM implementation, closed forms are derived for the convolution operator and the ADMM update of the full rank-one matrix. 
Numerical tests are performed to compare ROCO with four benchmark algorithms based on bilinear Lasso. 
Results demonstrate substantial improvement in recovery accuracy and comparable runtime of ROCO compared with benchmark algorithms. 


\clearpage

\bibliographystyle{ICASSPBibStyle}
\bibliography{References.bib}

\end{document}

%% file: preamble.tex
\usepackage{float}
\usepackage{mathrsfs}
\usepackage{mathtools}
\usepackage{bm}
\usepackage{bbm}
\usepackage{amsmath}
\usepackage{amsthm}
\usepackage{amssymb}
\usepackage{graphicx}
\usepackage{hyperref}

\usepackage{lipsum}

\usepackage{tabularx,booktabs}
\newcolumntype{C}[1]{>{\centering\arraybackslash}p{#1}}
\usepackage{array,multirow}

\makeatletter

\floatstyle{ruled}
\newfloat{algorithm}{tbp}{loa}
\providecommand{\algorithmname}{Algorithm}
\floatname{algorithm}{\protect\algorithmname}

\theoremstyle{plain}

\usepackage{thmtools}

\declaretheoremstyle[
  bodyfont=\normalfont\itshape,
  headformat=\NAME\NUMBER  
]{nospacetheorem}

\usepackage{filecontents} 
\theoremstyle{definition}

\@ifundefined{showcaptionsetup}{}{%
 \PassOptionsToPackage{caption=false}{subfig}}
\usepackage{subfig}
\makeatother

\providecommand{\lemmaname}{Lemma}
\providecommand{\propositionname}{Proposition}
\providecommand{\theoremname}{Theorem}